\documentclass[12pt]{article}

\usepackage[utf8]{inputenc}
\usepackage{amsmath,amssymb,mathtools,bm}
\usepackage[normalem]{ulem}
\usepackage{xcolor,graphicx,flafter}
\graphicspath{{figures/}{output/figures/}}
\usepackage{hyperref}
\hypersetup{
  unicode=true,
  colorlinks=true,
  linkcolor=blue!60!black,  
  citecolor=red!50!black,  
  urlcolor=blue!50!black,    
  pdftitle={Canonical Commutation Relations and Poisson Brackets with Non-locality in Time},
  pdfauthor={Yasu Gen, Hiroaki Matsunaga, Jojiro Totsuka}
}
\allowdisplaybreaks

\numberwithin{equation}{section}
\numberwithin{figure}{section}
\numberwithin{table}{section}

\newcommand{\dd}{\mathrm d}
\newcommand{\ii}{\mathrm i}
\newcommand{\ee}{\mathrm e}
\newcommand{\ad}{a^{\dagger}}
\newcommand{\Om}{\Omega}
\newcommand{\Lop}{\mathcal L}
\newcommand{\avg}[1]{\left\langle #1\right\rangle}
\newcommand{\PB}[2]{\left\{#1,#2\right\}}

\begin{document}

\baselineskip=17pt

\hypersetup{pageanchor=false}

\begin{titlepage}
\rightline{\tt KUNS-3139}

\begin{center}
\vskip 2.5cm
{\Large \bf {Canonical Commutation Relations and Poisson Brackets with Non-locality in Time}}

\vskip 1.0cm
{\large Yasu Gen$^{a,1}$, Hiroaki Matsunaga$^{b,2}$, Jojiro Totsuka$^{a,3}$}
\vskip 1.0cm

$ ^a${\it {Department of Physics, Kyoto University,}}\\
{\it {Kitashirakawa, Kyoto 606-8502, Japan}}\\

$ ^b${\it {Osaka Metropolitan University College of Technology, \\ Saiwai-cho, Neyagawa, Osaka 572-8572, Japan}}\\

$ ^1$yasugen@gauge.scphys.kyoto-u.ac.jp\\
$ ^2$via.string.field@gmail.com\\
$ ^3$georgeyo1227@gmail.com\\
\vspace{2mm}

\vskip 1.0cm

{\bf Abstract}
\end{center}
\noindent
We study the canonical structure of a string-inspired model in which the kinetic term couples fields at times separated by a finite shift $\sigma$,
with a coupling $\lambda$ to a fixed background. 
From the boundary term in the variation of the action, 
we construct a conserved symplectic form and derive the positive-time
Poisson bracket.
Using the Schwinger--Dyson relation,
we show that at positive time separation $s$ the commutator equals $\ii\hbar$ times this bracket.
The canonical normalization $\hbar$ appears as a jump of the commutator expectation value at $s=\sigma$, 
which is equal time for shifted variables with a local kinetic term; for $0<s<\sigma$ the commutator vanishes for every $\lambda>0$.
Hence $\lambda\to0$ at fixed $\sigma$ does not reproduce the free commutator on this interval,
whereas $\sigma\to0$ at fixed $\lambda$ recovers the local commutator at every fixed $s>0$.

\end{titlepage}
\hypersetup{pageanchor=true}

\tableofcontents

\section{Introduction}
\label{sec:introduction}

In string field theory (SFT), the extended nature of strings gives rise
to interactions containing infinitely many derivatives of the component
fields~\cite{EliezerWoodard,MoellerZwiebach}.
Since these include time derivatives, defining the solution space
and its canonical structure requires care.
For nondegenerate Lagrangians with a finite number of higher time
derivatives, Ostrogradsky's theorem implies an unbounded
Hamiltonian~\cite{WoodardOstrogradsky}.
This finite-order result does not directly settle the
infinite-derivative case, where the admissible solutions and
required initial data must be analyzed
separately~\cite{BarnabyKamran,ErbinFiratZwiebach}.

In theories with non-locality in time, the equations of motion (EOM)
couple fields at different times, and the symplectic form can involve
field variations over a time interval rather than at a single
time~\cite{LlosaVives}.
The equal-time canonical structure familiar from local theories
therefore need not apply directly.
String-inspired non-local scalar models and models with finite
time shifts provide concrete settings in which to examine this
issue~\cite{ErbinFiratZwiebach,Towards}.

On the quantum side, time ordering introduces further subtleties
in relating path-integral and operator formulations of theories
with non-locality in time~\cite{Fujikawa,LiaoSibold}.
In local theories,
the operator EOM and the contact terms determined by the canonical
commutation relations (CCR) together reproduce the
Schwinger--Dyson (SD) relation for time-ordered correlators.
With time shifts, the derivative and delayed terms can involve
different operator orderings, so separating the EOM and
commutator contributions requires care.
In the quantum Batalin--Vilkovisky (BV) formulation, the combination
of an EOM insertion and its quantum contact term is exact with
respect to the quantum BV differential $Q_\hbar$.
Expectation values on the cohomology therefore obey the SD
relation~\cite{FredenhagenRejzner,MasudaMatsunaga}.
In an operator realization, the EOM and commutator contributions
must jointly satisfy this relation~\cite{MatsunagaBVCCR}.

We study the finite-shift model of Ref.~\cite{Towards},
\begin{equation}
 S[a,\ad]=\int_{\mathbb R}\dd t\,
 [\ii\ad(t-\ell)\dot a(t+\ell)+\lambda b(t)a(t)\ad(t)]\,,
 \label{eq:action}
\end{equation}
where $\ell=\sigma/2>0$, $\lambda$ is the coupling to a fixed background
$b(t)$, and $a$ and $\ad$ are independent complex fields.
The model arises from exponential non-locality, which becomes a
translation along one light-cone direction after a Fourier expansion
along the other; the string-length parameter is analytically continued
so that the shift is real~\cite{Towards}.
For $b=1$ and $\lambda>0$, forward evolution according to
$\dot a(t)=\ii\lambda a(t-\sigma)$ requires a history over an interval
of length $\sigma$ and admits infinitely many characteristic modes,
whereas at $\lambda=0$ the field is a single constant.
The solution space on which the canonical structure is built thus
changes at zero coupling.

We address two questions. First, under specified conditions on the
correlators, does the commutator obtained from the SD relation agree
with $\ii\hbar$ times the Poisson bracket derived independently from the
classical action? Second, how does this correspondence behave when the
coupling or the time shift is sent to zero?

Perturbative reductions of higher-derivative and non-local theories
can restrict the solution space to a sector with a smooth limit
as the expansion parameter
vanishes~\cite{ErbinFiratZwiebach,Simon}.
Whether such a restriction preserves the brackets of the
observables under consideration must be checked.
Given a choice of vacuum and involution, Ref.~\cite{Towards}
constructs commutators that reproduce the path-integral two-point
function to all orders in the background coupling, and notes the
change in phase space at zero coupling and the failure of canonical
perturbation theory based on local free operators.

We complement the construction in Ref.~\cite{Towards} by deriving the classical bracket
independently from the action and identifying explicit correlator
conditions under which the two agree.
Our prescribed symmetric mode sum retains all characteristic modes
and allows us to quantify the contribution lost when only the
principal mode is kept; see
Eqs.~\eqref{eq:principal-and-remainder}
and \eqref{eq:coupling-mode-cancellation}.
On the classical side, we use covariant phase space (CPS)~\cite{CrnkovicWitten,Zuckerman,LeeWald}(see~\cite{Gieres:2021ekc} for review): the boundary
contribution to the variation of the action determines a symplectic
form on its solution space.
A Hamiltonian formalism for non-local Lagrangians was developed by
Llosa and Vives~\cite{LlosaVives} and extended in
Refs.~\cite{GomisKamimuraLlosa,HerediaLlosaMechanics,HerediaLlosaFields};
related constructions for non-locality of finite temporal extent and
their limitations appear in Refs.~\cite{WoodardFinite,LlosaComment}.
CPS methods for non-local theories have been developed in
Refs.~\cite{BernardesErlerFirat,AliStettinger} and applied in open
string field theory to rolling tachyons~\cite{ChoMazelYin,BEFPartI}
and moving lump solutions~\cite{BEFPartII}.
On the quantum side, we use the BV form of the SD relation described above.

Our main results are as follows.
\vspace{-2mm}
\begin{itemize}
\item
Varying the action on a finite time interval gives the conserved
symplectic form~\eqref{eq:omega}.
For $b=1$, its characteristic-mode decomposition and the summation
prescription of Appendix~\ref{app:spectral} give the classical Poisson bracket
\begin{equation}
 P_{\sigma,\lambda}(s)
 :=\PB{a(t)}{\ad(t')},
 \qquad s:=t-t'>0.
 \label{eq:P-definition-intro}
\end{equation}
Its explicit expression is given in Eq.~\eqref{eq:root-sum}.
\vspace{-3mm}
\item
The commutator expectation value
$B_{\sigma,\lambda}(s)=\avg{[\hat a(s),\hat a^\dagger(0)]}$ obeys
$B_{\sigma,\lambda}(s)=\ii\hbar P_{\sigma,\lambda}(s)$ for $s>0$,
under the evolution and continuity conditions specified in
Section~\ref{subsec:SD-kernel}.
The vacuum prescription of Ref.~\cite{Towards} realizes these conditions
order by order by making the reverse-ordered correlator vanish.
Under these conditions, $B$ jumps by $\hbar$ at $s=\sigma$ and vanishes
on $0<s<\sigma$.
\vspace{-3mm}
\item
At fixed $\sigma>0$, the additional modes retain a finite collective
contribution that cancels the principal mode on $0<s<\sigma$, even as
$\lambda\to0^+$, Eq.~\eqref{eq:coupling-mode-cancellation}.
By contrast, $\sigma\to0^+$ at fixed coupling recovers the local
commutator at every fixed $s>0$; the convergence is nonuniform near
equal time, Eq.~\eqref{eq:noncommuting-local-limits}.
\end{itemize}
\vspace{-2mm}
These results give a quantitative account of the singular
zero-coupling limit: modes absent from the strictly free theory
retain a finite collective contribution to the bracket.
Thus, in this model, weak coupling at a fixed time shift does not
justify replacing the full canonical structure by the free one.
The positive-time agreement between the independently derived
Poisson bracket and commutator expectation value,
shown in Eq.~\eqref{eq:correspondence}, establishes the
compatibility of the prescribed classical mode sum with
the conditions on quantum correlators stated in
Section~\ref{subsec:SD-kernel}.
It provides an explicit test of how a symplectic structure
derived from a non-local action is reflected in quantum
correlators, and motivates extending the comparison to the
non-local CPS constructions used in
SFT~\cite{BernardesErlerFirat,BEFPartI,BEFPartII}.

Section~\ref{sec:cps} derives the symplectic form and the Poisson
bracket, and Section~\ref{sec:sd} derives the commutator from the SD
relation. Section~\ref{sec:comparison} compares the two and examines
the zero-coupling and zero-delay limits, and
Section~\ref{sec:discussion} concludes.
Appendices~\ref{app:spectral} and~\ref{app:distributional-extension}
evaluate the mode sum and construct a distributional extension to
nonpositive separation.

\section{Classical Poisson brackets from covariant phase space}
\label{sec:cps}

\subsection{The model and conventions}
\label{sec:model}

The finite-shift model introduced above allows an explicit comparison of
the classical Poisson structure with the quantum SD relation. We now fix
the field conventions and the meaning of the time cutoff.

We consider the action~\eqref{eq:action} with $\sigma>0$ and $\ell=\sigma/2$%
\footnote{We omit the tildes used for the field variables in Ref.~\cite{Towards}.}. The background $b$ is fixed, so $\delta b=0$. We regard the action as a
functional of independent complex fields $a$ and $\ad$, without imposing the
usual complex-conjugation condition $\ad(t)=a(t)^*$.

Shifting the time arguments of the two independent fields in opposite
directions, we define
\begin{equation}
 A(u):=a(u+\ell),\qquad
 \bar A(u):=\ad(u-\ell)\,.
 \label{eq:local-kinetic-variables}
\end{equation}
The same action then takes the form
\begin{equation}
 S=\int_{\mathbb R}\dd u\,
 \left[
 \ii\bar A(u)\dot A(u)
 +\lambda b(u)A(u-\ell)\bar A(u+\ell)
 \right]\,.
 \label{eq:local-kinetic-action}
\end{equation}
At $\lambda=0$, it therefore reduces to a local free action for any $\sigma$; at $\sigma=0$, the coupling term also becomes local. 
This change of variables moves the non-locality from the kinetic term to the background coupling term.

After integration by parts, the functional derivatives of the action are
\begin{align}
 E_{\ad}(t)&:=\frac{\delta S}{\delta\ad(t)}=\ii\dot a(t+\sigma)+\lambda b(t)a(t),
 \label{eq:E-ad}\\
 E_a(t)&:=\frac{\delta S}{\delta a(t)}=-\ii\dot\ad(t-\sigma)+\lambda b(t)\ad(t)\,.
 \label{eq:E-a}
\end{align}
Even for real $\lambda$ and $b$, these two expressions are not related by
ordinary complex conjugation: the derivative term in the equation for $a$
contains $t+\sigma$, whereas that for $\ad$ contains $t-\sigma$.
Relabeling the time arguments gives
\begin{align}
 \dot a(t)&=\ii\lambda b(t-\sigma)a(t-\sigma),
 \label{eq:eom-a}\\
 \dot\ad(t)&=-\ii\lambda b(t+\sigma)\ad(t+\sigma)\,.
 \label{eq:eom-ad}
\end{align}
The first equation involves a past time argument, while the second involves
a future one.

Throughout this paper, field configurations are defined on the entire time
axis. The time cutoff restricts only the integration domain of the action.
We first vary the action on a finite interval $[\tau_0,\tau]$ and then restrict
to classical solutions satisfying the EOM on the full time axis. This
distinction is necessary because changes of variables in the kinetic term
shift the upper integration endpoint to $\tau-\ell$ or $\tau+\ell$, while the
endpoint of the background coupling remains at $\tau$. Before using the EOM,
we must bring the integration domains to a common interval and retain their
differences as boundary contributions. The next subsection carries out this
procedure explicitly.

\subsection{Symplectic potential and symplectic form}

In a local theory, the symplectic form can be read off from the endpoint term
in the variation of the action. We follow the same strategy here, including
the contributions from finite intervals that straddle the endpoint.
The symbol $\delta$ denotes variation on the space of field configurations
and is distinct from a time derivative. We hold $\lambda$, $\sigma$, $b$,
and the integration endpoints fixed, and vary $a$ and $\ad$ independently.

Consider the action on a finite interval,
\begin{equation}
 S_{[\tau_0,\tau]}=\int_{\tau_0}^{\tau}\dd t\,\,
 [\ii\ad(t-\ell)\dot a(t+\ell)
   +\lambda b(t)a(t)\ad(t)]\,.
 \label{eq:finite-cut-action}
\end{equation}
On a finite time interval, integration by parts and time shifts
generate boundary contributions, as shown below.
Analogous contributions appear as a violation of cyclicity
in string field theory~\cite{GenMatsunagaTotsuka}.
Integrating the kinetic term by parts and using the argument of each field
variation as the integration variable, we find
\begin{align}
 \delta S_{[\tau_0,\tau]}={}&
 [\ii\ad(t-\ell)\delta a(t+\ell)]_{\tau_0}^{\tau}
 +\ii\int_{\tau_0-\ell}^{\tau-\ell}\dd u\,
 \dot a(u+\sigma)\delta\ad(u)
 \nonumber\\
 &-\ii\int_{\tau_0+\ell}^{\tau+\ell}\dd u\,
 \dot\ad(u-\sigma)\delta a(u)
 +\lambda\int_{\tau_0}^{\tau}\dd u\,b(u)
 [\ad(u)\delta a(u)+a(u)\delta\ad(u)]\,.
 \label{eq:shifted-variation}
\end{align}
Bringing the three integrals to the common interval $[\tau_0,\tau]$ and
retaining the differences gives
\begin{equation}
 \delta S_{[\tau_0,\tau]}
 =\int_{\tau_0}^{\tau}\dd u\,
 [E_a(u)\delta a(u)+E_{\ad}(u)\delta\ad(u)]
 +\vartheta_\tau-\vartheta_{\tau_0}\,.
 \label{eq:finite-variation}
\end{equation}
The ``boundary'' one-form, or symplectic potential, is
\begin{align}
 \vartheta_\tau={}&
 \ii\ad(\tau-\ell)\delta a(\tau+\ell)
 -\ii\int_\tau^{\tau+\ell}\dd u\,
 \dot\ad(u-\sigma)\delta a(u)
 -\ii\int_{\tau-\ell}^{\tau}\dd u\,
 \dot a(u+\sigma)\delta\ad(u)\,.
 \label{eq:theta}
\end{align}
The first term is the usual endpoint contribution from integration by parts.
The second and third terms arise from the differences between the integration
domains of the kinetic and background coupling terms.

Here $\delta$ denotes the exterior derivative on field space.
We define $\Om_\tau$ by restricting $\delta\vartheta_\tau$
to the solution space.
Variations within this space satisfy the linearized EOM:
\begin{equation}
 \delta\dot a(u+\sigma)=\ii\lambda b(u)\delta a(u),\qquad
 \delta\dot\ad(u-\sigma)=-\ii\lambda b(u)\delta\ad(u)\,.
 \label{eq:linearized-eom}
\end{equation}
The exterior derivatives of the two integrals in Eq.~\eqref{eq:theta}
therefore combine into
$-\lambda\int_{\tau-\ell}^{\tau+\ell}b\,\delta\ad\wedge\delta a$.
The symplectic form on the solution space is thus
\begin{equation}
 \Om_\tau=
 \ii\delta\ad(\tau-\ell)\wedge\delta a(\tau+\ell)
 -\lambda\int_{\tau-\ell}^{\tau+\ell}\dd u\,
 b(u)\delta\ad(u)\wedge\delta a(u)\,.
 \label{eq:omega}
\end{equation}
This calculation gives a concrete realization, for the present model, of the integral construction of symplectic two-forms for non-local actions~\cite{HerediaLlosaMechanics,HerediaLlosaFields}.
One can verify that the Llosa--Vives canonical one-form~\cite{LlosaVives},
expressed in the same field variables and evaluated with the same
time cutoff, agrees with Eq.~\eqref{eq:theta} when restricted to the solution space.
Taking its exterior derivative then reproduces Eq.~\eqref{eq:omega}.

Conservation of Eq.~\eqref{eq:omega} follows directly by differentiating
with respect to the cutoff time. Using the linearized EOM, we find for the
endpoint term
\begin{align}
 \frac{\dd}{\dd\tau}
 [\ii\delta\ad(\tau-\ell)\wedge\delta a(\tau+\ell)]
 ={}&
 \lambda b(\tau+\ell)
 \delta\ad(\tau+\ell)\wedge\delta a(\tau+\ell)
 \nonumber\\
 &-\lambda b(\tau-\ell)
 \delta\ad(\tau-\ell)\wedge\delta a(\tau-\ell)\,.
 \label{eq:boundary-derivative}
\end{align}
This is canceled by the $\tau$-derivative of the integral term, which acts
on its endpoints. Since
$\Om=\delta\vartheta$, the form is also closed:
\begin{equation}
 \frac{\dd\Om_\tau}{\dd\tau}=0,\qquad
 \delta\Om_\tau=0\,.
 \label{eq:conserved}
\end{equation}
The endpoint contribution, given by the first term in Eq.~\eqref{eq:omega},
is generally not conserved by itself.
Its time derivative is canceled by that of the second term,
which arises from the shifted integration intervals.
Both terms are therefore needed for the conservation law in Eq.~\eqref{eq:conserved}.

Evaluating this symplectic form requires only the tangent variations on the interval
$I_\tau=[\tau-\ell,\tau+\ell]$, whose width is $\sigma$. The single-time slice of
a local theory is thus replaced by a time interval of finite thickness.

For configurations and variations with a smooth, bounded local limit, the
integral term vanishes as $\sigma\to0$, while the endpoint term approaches
$\ii\delta\ad(\tau)\wedge\delta a(\tau)$. However, the set of solution
modes also changes with $\sigma$. The limit of the symplectic form alone
therefore does not determine the limit of its inverse bracket. We examine
this issue explicitly in Section~\ref{sec:comparison}.

\subsection{Poisson brackets}
\label{sec:poisson}

We now set $b(t)=1$ and take $\lambda>0$. 
Using the time-shift operator
$\ee^{-\sigma\partial_t}a(t)=a(t-\sigma)$,
we write the equation for $a$ as
\begin{equation}
 \Lop a(t)=0,\qquad
 \Lop=\ii\partial_t+\lambda\ee^{-\sigma\partial_t}\,.
 \label{eq:Lop}
\end{equation}

For a mode $a(t)\propto\ee^{\ii kt}$, the characteristic equation and its
roots are
\begin{equation}
 k=\lambda\ee^{-\ii k\sigma},\qquad
 k_n=\frac{W_n(\ii\lambda\sigma)}{\ii\sigma},
 \qquad n\in\mathbb Z\,.
 \label{eq:lambert}
\end{equation}
Here $W_n$ denotes the $n$th branch of the Lambert $W$ function, satisfying
$W_n(z)\ee^{W_n(z)}=z$~\cite{DLMF,Cohen}. 
We use the formal characteristic-mode expansions
\begin{equation}
 a(t)=\sum_n\alpha_n\ee^{\ii k_nt},\qquad
 \ad(t)=\sum_n\beta_n\ee^{-\ii k_nt}\,.
 \label{eq:modes}
\end{equation}
The opposite signs in the exponents allow the same set of roots to be used
for both fields. The coefficients $\alpha_n$ and $\beta_n$ are independent.

Substituting these expansions into Eq.~\eqref{eq:omega}, we find that the terms
$\delta\beta_m\wedge\delta\alpha_n$ with $m\neq n$ cancel%
\footnote{
For $m\neq n$, the coefficient of
$\delta\beta_m\wedge\delta\alpha_n$ is
\[
 \ii\ee^{\ii(k_n-k_m)\tau+\ii(k_n+k_m)\ell}
 \left[
  1-\lambda
  \frac{\ee^{-\ii k_n\sigma}-\ee^{-\ii k_m\sigma}}
       {k_n-k_m}
 \right].
\]
This vanishes by the characteristic equation
$k_j=\lambda\ee^{-\ii k_j\sigma}$.
}
, leaving
\begin{equation}
 \Om=\sum_n C_n\delta\beta_n\wedge\delta\alpha_n,\qquad
 C_n=\ii\ee^{\ii k_n\sigma}(1+\ii k_n\sigma)\,.
 \label{eq:omega-modes}
\end{equation}
The factor $1+\ii k_n\sigma$ results from combining the endpoint
contribution $\ii\ee^{\ii k_n\sigma}$ with the interval contribution
$-\lambda\sigma$. In terms of the characteristic function
$D(k)=k-\lambda\ee^{-\ii k\sigma}$, the coefficient can also be written
as $C_n=\ii\ee^{\ii k_n\sigma}D'(k_n)$.
For real positive $\lambda$ and $\sigma$, every $C_n$ is nonzero. 
On each finite-mode subspace, the Poisson bracket is defined by the inverse of the symplectic form.
For the convention $\Om_N=\sum_{n=-N}^{N}C_n\delta\beta_n\wedge\delta\alpha_n$,
this gives
\begin{align}
     \PB{F}{G}_N
 :=
 \sum_{n=-N}^{N}\frac{1}{C_n}
 \left(
 \frac{\partial F}{\partial\alpha_n}
 \frac{\partial G}{\partial\beta_n}
 -
 \frac{\partial F}{\partial\beta_n}
 \frac{\partial G}{\partial\alpha_n}
 \right).
\end{align}
In particular, the mode coefficients satisfy%
\footnote{
Here the mode brackets are defined on the finite-mode subspaces with
$|n|\leq N$, on which $\Om$ restricts to the nondegenerate
form~\eqref{eq:omega-finite}, and the field bracket below is defined by
the $N\to\infty$ limit~\eqref{eq:root-sum-limit} of
Appendix~\ref{app:spectral}.
}
\begin{equation}
 \PB{\alpha_n}{\beta_m}=\frac{\delta_{nm}}{C_n},\qquad
 \PB{\alpha_n}{\alpha_m}=\PB{\beta_n}{\beta_m}=0\,.
 \label{eq:mode-bracket}
\end{equation}

Related mode-bracket constructions for non-local oscillators appear in Ref.~\cite{HerediaLlosaMechanics}. 
For positive time separation $s=t-t'>0$, the bracket is given by 
\begin{equation}
 P_{\sigma,\lambda}(s):=\PB{a(t)}{\ad(t')}
 =\sum_{n}^{\mathrm{sym}}
 \frac{\ee^{\ii k_n(s-\sigma)}}{\ii(1+\ii k_n\sigma)}\,.
 \label{eq:root-sum}
\end{equation}
Here the summation is defined in Appendix~\ref{app:spectral}, where its construction from the symplectic form and its convergence for $s>0$ are established.
Evaluating this sum gives
\begin{equation}
 P_{\sigma,\lambda}(s)=
 -\ii\sum_{m=0}^{\infty}\frac{(\ii\lambda)^m}{m!}
 [s-(m+1)\sigma]^m\Theta(s-(m+1)\sigma),
 \qquad s>0\,.
 \label{eq:P-positive}
\end{equation}
Here $\Theta$ is the Heaviside function, 
with $\Theta(0)=1/2$, 
and the $m=0$ term is understood as $-\ii\Theta(s-\sigma)$. 
For $\sigma>0$, this gives
\begin{equation}
 P_{\sigma,\lambda}(s)=0\quad(0<s<\sigma),\qquad
 P_{\sigma,\lambda}(s)=-\ii\quad(\sigma<s<2\sigma)\,.
 \label{eq:P-first-intervals}
\end{equation}
At fixed $\sigma>0$, the vanishing of the bracket on $0<s<\sigma$
persists as $\lambda\to0^+$, whereas setting $\lambda=0$ gives
$P_{\mathrm{free}}(s)=-\ii$.
Thus, an arbitrarily small coupling does not justify replacing the
bracket by that of the strictly free theory.
Section~\ref{sec:comparison} explains this discrepancy in terms of the
additional characteristic modes and shows that the zero-delay limit
instead recovers the local bracket at fixed $\lambda>0$ and $s>0$.

\section{SD relation and the quantum commutator}
\label{sec:sd}

\subsection{Quantum BV interpretation of the SD relation}
\label{subsec:quantum-cohomology}

We use the quantum BV differential to make explicit the contact
terms that accompany EOM insertions in quantum correlators.
For this model without gauge symmetry, let $\phi^I=(a,\ad)$ be independent bosonic fields of degree zero,
and associate to each field a Grassmann-odd antifield
$\phi_I^*$ of degree $-1$.
With $E_I(u)=\delta S/\delta\phi^I(u)$ as in Eqs.~\eqref{eq:E-ad} and \eqref{eq:E-a}, we define
\begin{align}
 Q_\hbar&=Q_{\rm cl}-\ii\hbar\Delta,
 \label{eq:quantum-BV-differential}\\
 Q_{\rm cl}&=\sum_I\int\dd u\,E_I(u)
 \frac{\delta^L}{\delta\phi_I^*(u)},
 &
 \Delta&=\sum_I(-1)^{\epsilon_I}\int\dd u\,
 \frac{\delta^L}{\delta\phi^I(u)}
 \frac{\delta^L}{\delta\phi_I^*(u)}\,,
 \label{eq:BV-components}
\end{align}
with conventions appropriate to the weight $\ee^{\ii S/\hbar}$.
Here $\epsilon_I$ is the Grassmann parity of $\phi^I$,
and $\delta^L$ denotes the left functional derivative with respect
to either a field or an antifield.
In the present model $\epsilon_I=0$ for all $I$, so the sign factors equal one.
The operators $E_I-\ii\hbar\delta/\delta\phi^I$ commute with one another,
while the antifield derivatives anticommute, giving $Q_\hbar^2=0$.
For the quantum BV description of time-ordered products and
perturbative path integration, 
see Refs.~\cite{FredenhagenRejzner,MasudaMatsunaga}.
A related construction of correlators satisfying the SD equations
uses quantum $A_\infty$ algebras~\cite{Okawa}.

An antifield-independent functional $F$ is $Q_\hbar$-closed, and
\begin{equation}
 Q_\hbar(\phi_I^*(u)F)
 =E_I(u)F-\ii\hbar\frac{\delta F}{\delta\phi^I(u)}\,.
 \label{eq:Q-exact-SD}
\end{equation}
Thus, the combination of the EOM insertion and the contact term on the
right-hand side is $Q_\hbar$-exact, and hence trivial in cohomology. For the expectation value
to define a linear functional on this cohomology, it must satisfy%
\footnote{
Here $\avg{\cdots}$ of fields without hats denotes a path-integral expectation value, while hatted fields denote operators.}
\begin{equation}
 \avg{E_I(u)F}
 =\ii\hbar
 \avg{\frac{\delta F}{\delta\phi^I(u)}}\,.
 \label{eq:cohomological-SD}
\end{equation}
This is precisely the SD identity obtained by functional integration by
parts. In particular, for $F=\phi^J(v)$,
\begin{equation}
 Q_\hbar(\phi_I^*(u)\phi^J(v))
 =E_I(u)\phi^J(v)-\ii\hbar\delta_I{}^J\delta(u-v).
 \label{eq:Q-exact-two-point}
\end{equation}
In the classical limit, $Q_{\rm cl}\phi_I^*=E_I$ implements the quotient by
the EOM. In the quantum theory, however, the exactness of $E_I$ does not
imply that $E_I F$ is exact. Setting the EOM to zero unconditionally inside
a product would discard the contact term in Eq.~\eqref{eq:Q-exact-SD}.
In taking $Q_\hbar$-cohomology, the EOM insertion must be treated together with its quantum contact term. 
In the operator formulation, this single
relation is realized by a combination of the EOM and the canonical
commutation relations (CCR)~\cite{MatsunagaBVCCR}.

Consider first the local theory with $b=1$ and $\sigma=0$. Define
$G_{\rm loc}(t,t')=\avg{T\hat a(t)\hat a^\dagger(t')}$ and
$\widehat E_{\rm loc}(t)=\ii\partial_t\hat a(t)+\lambda\hat a(t)$.
Differentiating the ordinary time-ordered product from outside the
time-ordering symbol gives
\begin{align}
 \ii\hbar\delta(t-t')
 &=(\ii\partial_t+\lambda)G_{\rm loc}(t,t')
 =\avg{T[\widehat E_{\rm loc}(t)\hat a^\dagger(t')]}
 +\ii\delta(t-t')\avg{[\hat a(t),\hat a^\dagger(t)]}\,.
 \label{eq:local-EOM-CCR-balance}
\end{align}
When the field operators satisfy the EOM, the EOM insertion on the
right-hand side vanishes. Comparing the coefficients of the contact term then yields $\avg{[\hat a(t),\hat a^\dagger(t)]}=\hbar$.
If the commutator is a $c$-number, this is the usual CCR.

Equation~\eqref{eq:local-EOM-CCR-balance} shows how the EOM and CCR
jointly realize the local SD relation.
The BV formulation gives this relation a cohomological interpretation,
but does not by itself determine the commutator.
For the non-local model, the next subsection supplements the SD
relation with explicit evolution and continuity conditions on the
correlators to determine the positive-time commutator.

\subsection{The SD relation on each interval and the commutator}
\label{subsec:SD-kernel}
We set $b=1$ and start from the functional-variation identity
\begin{equation}
 0=\int\mathcal D a\,\mathcal D\ad\,
 \frac{\delta}{\delta\ad(t-\sigma)}
 [\ad(t')\ee^{\ii S/\hbar}]\,.
 \label{eq:SD-variation}
\end{equation}
The derivatives of the insertion and of the action are
\begin{equation}
 \frac{\delta\ad(t')}{\delta\ad(t-\sigma)}
 =\delta(t-t'-\sigma),\qquad
 \frac{\delta S}{\delta\ad(t-\sigma)}
 =\ii\dot a(t)+\lambda a(t-\sigma)\,.
 \label{eq:SD-functional-derivatives}
\end{equation}
The path-integral two-point function
$G(t,t'):=\avg{a(t)\ad(t')}$ therefore satisfies
\begin{equation}
 \ii\partial_tG(t,t')+\lambda G(t-\sigma,t')
 =\ii\hbar\delta(t-t'-\sigma)\,.
 \label{eq:SD}
\end{equation}
The contact source is supported at $t=t'+\sigma$ because the variation
is taken with respect to the field at time $t-\sigma$.

We set $t'=0$ and write $s=t$.
Since the SD equation~\eqref{eq:SD} has time-independent coefficients
for $b=1$, the same argument holds for any $t'$, with $s=t-t'$.
We examine the conditions under which the
path-integral two-point function can be realized by ordinary time
ordering with respect to the original time labels:
\begin{equation}
 G(s)=
 \Theta(s)\avg{\hat a(s)\hat a^\dagger(0)}
 +\Theta(-s)\avg{\hat a^\dagger(0)\hat a(s)}\,.
 \label{eq:ordering}
\end{equation}
The distinction between the $T^*$ product and the ordinary $T$ product in
theories with higher time derivatives is discussed in Ref.~\cite{Fujikawa};
issues of time ordering for interactions non-local in time are discussed
in Ref.~\cite{LiaoSibold}. In the SD relation, the derivative acts on the
entire function $G$.

The operator ordering in the derivative term is set by the sign of $s$,
whereas that in the delay term is set by the sign of $s-\sigma$.
Away from the contact points, the SD equation thus
splits into three regions:
\begin{equation}
 0=
 \begin{cases}
 \ii\partial_s\avg{\hat a^\dagger(0)\hat a(s)}
 +\lambda\avg{\hat a^\dagger(0)\hat a(s-\sigma)},
 &s<0,\\[4pt]
 \ii\partial_s\avg{\hat a(s)\hat a^\dagger(0)}
 +\lambda\avg{\hat a^\dagger(0)\hat a(s-\sigma)},
 &0<s<\sigma,\\[4pt]
 \ii\partial_s\avg{\hat a(s)\hat a^\dagger(0)}
 +\lambda\avg{\hat a(s-\sigma)\hat a^\dagger(0)},
 &s>\sigma
 \end{cases}\,.
 \label{eq:SD-three-regions}
\end{equation}
The operator orderings of the two terms agree in the outer regions and
differ in the middle one. In what follows, equations on individual
intervals are understood on open intervals excluding the matching points
$s=n\sigma$, with $n=0,1,\ldots$. The ordering mismatch leaves a
contribution from
\begin{equation}
 B_{\sigma,\lambda}(s)=\avg{[\hat a(s),\hat a^\dagger(0)]}\,.
 \label{eq:Wightman}
\end{equation}
We refer to $B$ as the commutator below, with the understanding that
it is an expectation value. It also determines the operator commutator
when the latter is a $c$-number.

To determine $B$ for positive time separation, we assume that the reverse-ordered
two-point function satisfies the homogeneous EOM on the open intervals:
\begin{equation}
 \ii\partial_s\avg{\hat a^\dagger(0)\hat a(s)}
 +\lambda\avg{\hat a^\dagger(0)\hat a(s-\sigma)}=0,
 \qquad s>0,\quad s\neq n\sigma\,.
 \label{eq:ordered-EOM-condition}
\end{equation}
This condition is imposed on the chosen realization of the correlators;
it does not follow from the SD relation alone. The matching points
$n\sigma$, with $n=1,2,\ldots$, will be treated separately using the SD
relation. Subtracting this homogeneous equation from
Eq.~\eqref{eq:SD-three-regions} gives
\begin{equation}
 \begin{cases}
 B_{\sigma,\lambda}'(s)=0,
 &0<s<\sigma,\\[2pt]
 \ii B_{\sigma,\lambda}'(s)+\lambda B_{\sigma,\lambda}(s-\sigma)=0,
 &s>\sigma,\quad s\neq n\sigma\,.
 \end{cases}
 \label{eq:B-delay}
\end{equation}
Hence $B_{\sigma,\lambda}(s)=C$ on the first interval, with $C$ undetermined at this stage.
We next integrate the SD equation itself to determine the matching
conditions, rather than imposing the homogeneous EOM at the contact
points.

We assume that both orderings of the two-point function remain bounded
near the matching points. When Eq.~\eqref{eq:SD} is integrated over a
shrinking interval around $s_0$, the integrated delay term vanishes in
the limit, and we obtain
\begin{equation}
 G(s_0^+)-G(s_0^-)=
 \begin{cases}
 \hbar,&s_0=\sigma,\\
 0,&s_0\neq\sigma
 \end{cases}\,.
 \label{eq:G-jump}
\end{equation}
In particular, at $s=\sigma>0$,
\begin{equation}
 \avg{\hat a(\sigma^+)\hat a^\dagger(0)}
 -\avg{\hat a(\sigma^-)\hat a^\dagger(0)}
 =\hbar.
 \label{eq:direct-jump}
\end{equation}
Here the one-sided limits are taken on the two-point function.

To interpret this as a jump of the commutator, we additionally assume that
the reverse-ordered two-point function
$\avg{\hat a^\dagger(0)\hat a(s)}$ is continuous at the positive matching
points $s=n\sigma$, with $n\geq1$. Together with Eq.~\eqref{eq:G-jump},
this gives
\begin{align}
 B_{\sigma,\lambda}(s)&=C,
 &&0<s<\sigma,
 \label{eq:B-strip}\\
 B_{\sigma,\lambda}(\sigma^+)&=C+\hbar,
 &&
 \label{eq:B-normalization}\\
 B_{\sigma,\lambda}(n\sigma^+)&=B_{\sigma,\lambda}(n\sigma^-),
 &&n\geq2\,.
 \label{eq:B-later-matching}
\end{align}

Since $\sigma>0$, the SD equation has no contact source at the origin and
requires $G(0^+)=G(0^-)$. If the reverse-ordered two-point function is also
assumed to be continuous at $s=0$, $B_{\sigma,\lambda}(0^+)=0$, and hence
$C=0$. This continuity is an additional condition on the correlators.
Under it, $B$ vanishes on the first interval and jumps by $\hbar$ at
$s=\sigma$.
For the variables $A(u)=a(u+\ell)$ and $\bar A(v)=\ad(v-\ell)$ in Eq.~\eqref{eq:local-kinetic-variables}, the original separation is $s=u-v+\sigma$, so this commutator jump occurs at equal time $u=v$ in the variables whose kinetic term is local.
The jump in Eq.~\eqref{eq:direct-jump} means that the
relevant field matrix element is discontinuous; weak continuity of
$\hat a(s)$ on a common domain containing these states is not assumed.

We solve the interval equations with these matching conditions by expanding
in the coupling:
\begin{equation}
 B_{\sigma,\lambda}(s)=\sum_{m=0}^{\infty}\lambda^mB_m(s)\,.
 \label{eq:B-expansion}
\end{equation}
On the open intervals with $s>\sigma$,
\begin{equation}
 \ii B_0'(s)=0,\qquad
 \ii B_m'(s)+B_{m-1}(s-\sigma)=0
 \quad(m\geq1)\,.
 \label{eq:B-recursion}
\end{equation}
We first solve the equation on successive intervals.
On $\sigma<s<2\sigma$, the delayed term vanishes because
$B_{\sigma,\lambda}(s-\sigma)=0$.
Thus $B_{\sigma,\lambda}'(s)=0$, and the jump condition
$B_{\sigma,\lambda}(\sigma^+)=\hbar$ gives $B_{\sigma,\lambda}(s)=\hbar$.
On $2\sigma<s<3\sigma$, we have $B_{\sigma,\lambda}(s-\sigma)=\hbar$, so
$B_{\sigma,\lambda}'(s)=\ii\lambda\hbar$.
Using continuity at $s=2\sigma$, we obtain
\[
 B_{\sigma,\lambda}(s)=B_{\sigma,\lambda}(2\sigma)+\ii\lambda\hbar(s-2\sigma)
 =\hbar+\ii\lambda\hbar(s-2\sigma).
\]
On $3\sigma<s<4\sigma$, the preceding solution gives
\[
 B_{\sigma,\lambda}(s-\sigma)=\hbar+\ii\lambda\hbar(s-3\sigma),
 \qquad
 B_{\sigma,\lambda}'(s)=\ii\lambda\hbar-\lambda^2\hbar(s-3\sigma).
\]
$B_{\sigma,\lambda}(3\sigma)=\ldots$ fixes the next interval.
The first few intervals are therefore
\begin{equation}
 B_{\sigma,\lambda}(s)=
 \begin{cases}
 0,
 &0<s<\sigma,\\[2pt]
 \hbar,
 &\sigma<s<2\sigma,\\[2pt]
 \hbar+\ii\hbar\lambda(s-2\sigma),
 &2\sigma<s<3\sigma,\\[2pt]
 \hbar+\ii\hbar\lambda(s-2\sigma)
 -\dfrac{\hbar\lambda^2}{2}(s-3\sigma)^2,
 &3\sigma<s<4\sigma
 \end{cases}\,.
 \label{eq:B-intervals}
\end{equation}
Each successive interval introduces one further power of $\lambda$.
Continuing the recursion gives
\begin{equation}
 B_{\sigma,\lambda}(s)=
 \hbar\sum_{m=0}^{\infty}\frac{(\ii\lambda)^m}{m!}
 [s-(m+1)\sigma]^m\Theta(s-(m+1)\sigma),
 \qquad s>0\,.
 \label{eq:B-closed-form}
\end{equation}
The $m=0$ term is $\hbar\Theta(s-\sigma)$ and produces the jump at
$s=\sigma$. The SD relation does not fix the value at this single point.
For comparison with Eq.~\eqref{eq:P-positive}, we choose
$\Theta(0)=1/2$.
For $m\geq1$, the factor multiplying the derivative of the step function
vanishes at the matching point, while the remaining derivative cancels
the delay term from the preceding order in Eq.~\eqref{eq:B-recursion}.
Thus the expression satisfies the interval equations and the matching
conditions.
Imposing the conditions above at each $t'$, the same argument gives
$\avg{[\hat a(t),\hat a^\dagger(t')]}=B_{\sigma,\lambda}(t-t')$ for $t>t'$.

Starting instead from the SD relation for $\ad$, obtained by varying
$a(t'+\sigma)$, gives the same commutator~\eqref{eq:B-closed-form}.
The derivative and the continuity conditions then refer to the time of
$\ad$, and the resulting equations in $t'$ give the same function of $t-t'$.

For real $\sigma>0$, the vacuum prescription of Ref.~\cite{Towards}
provides a formal perturbative realization of these assumptions.
Its Eq.~(3.12) imposes $\hat a_0(t)|0\rangle=0$ for every $t$, and its
recursion relation (E.1) ensures that $\hat a(t)|0\rangle=0$ holds order
by order.
Thus $\langle0|\hat a^\dagger(0)\hat a(s)|0\rangle=0$, which satisfies
Eq.~\eqref{eq:ordered-EOM-condition} and the continuity conditions.
For $b=1$, its Eq.~(3.9), with $\hbar$ restored, gives
Eq.~\eqref{eq:B-closed-form} at positive separation away from the jump;
the value at $s=\sigma$ depends on the contact-point convention.

\section{Classical--quantum correspondence and limits}
\label{sec:comparison}

We now compare the classical Poisson bracket
$P_{\sigma,\lambda}$ obtained from the symplectic form with the
quantum commutator expectation value $B_{\sigma,\lambda}$.
Comparing Eqs.~\eqref{eq:P-positive} and \eqref{eq:B-closed-form}, we find
\begin{equation}
 B_{\sigma,\lambda}(s)=
 \ii\hbar P_{\sigma,\lambda}(s),
 \qquad s>0\,.
 \label{eq:correspondence}
\end{equation}
On the classical side, this relation uses the specified summation
prescription over all characteristic roots. On the quantum side, it uses
the homogeneous EOM in Eq.~\eqref{eq:ordered-EOM-condition} and the
continuity of the reverse-ordered two-point function at the matching
points, including the origin.

We next examine whether taking a parameter to zero in this commutator agrees with first setting that parameter to zero in the action and then constructing the canonical structure. The zero-coupling and zero-delay limits give different answers.
At fixed $\sigma>0$, the limit $\lambda \rightarrow 0^+$ does not recover the strictly free commutator on $0<s<\sigma$.
At fixed $\lambda>0$, the limit $\sigma \rightarrow 0^+$ does
recover the local commutator at every fixed $s>0$, but the convergence is nonuniform near equal time.

For reference, setting $\sigma=0$ in the action gives the local equation $\dot a=\ii\lambda a$ with the usual canonical normalization. Its commutator and zero-coupling limit are
\begin{equation}
 B_{\mathrm{loc},\lambda}(s)=\hbar\ee^{\ii\lambda s},
 \qquad
 \lim_{\lambda\to0^+}B_{\mathrm{loc},\lambda}(s)=\hbar\,.
 \label{eq:local-reference-commutator}
\end{equation}
Thus, in this local theory, taking the coupling to zero reproduces the free commutator continuously.

\subsection{Zero-coupling limit at fixed delay} 
Consider first the zero-coupling limit at fixed delay
$\sigma>0$ and fixed separation $s>0$.
At exactly $\lambda=0$, the EOM become
$\dot a={\dot a}^\dagger=0$, and the solution space and its canonical structure are
\begin{equation}
 a(t)=\alpha,\qquad
 \ad(t)=\beta,\qquad
 \Om=\ii\delta\beta\wedge\delta\alpha,\qquad
 P_{\mathrm{free}}(s)=-\ii\,.
 \label{eq:strict-free}
\end{equation}
Quantizing this single canonical pair gives $B_{\mathrm{free}}(s)=\hbar$ for any $\sigma$.

\begin{figure}[htbp]
 \centering
 \includegraphics[width=0.65\linewidth]{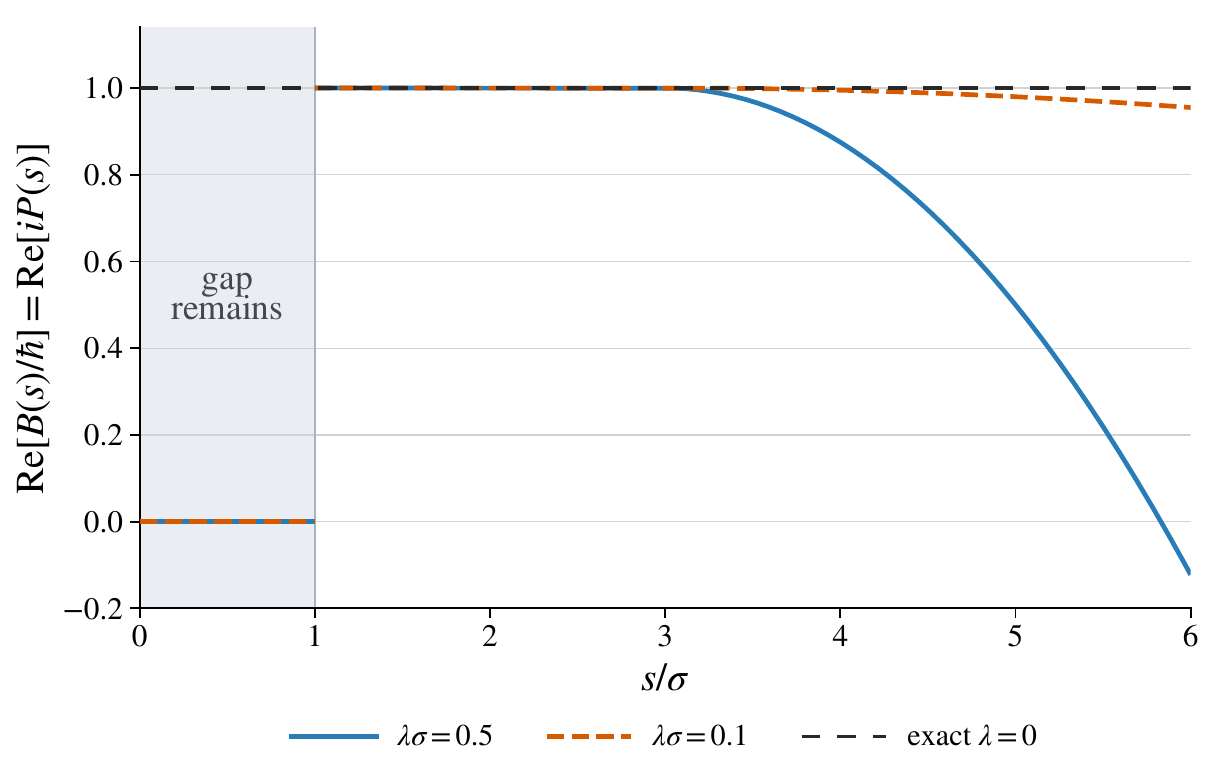}
 \caption{
 Comparison of the zero-coupling limit with the strictly free theory.
 The colored curves show the real part of the normalized commutator,
 $\operatorname{Re}[B_{\sigma,\lambda}(s)/\hbar]=\operatorname{Re}[\ii P_{\sigma,\lambda}(s)]$,
 for several values of $\lambda\sigma>0$. The black dashed line is the
 free theory defined at exactly $\lambda=0$. For every $\lambda\sigma>0$,
 the normalized commutator vanishes on the shaded interval $0<s<\sigma$,
 whereas the free-theory value is one. 
 }
 \label{fig:zero-limit}
\end{figure}

For $\lambda>0$, however, the full commutator is given by Eq.~\eqref{eq:B-closed-form}.
At fixed $s$ and $\sigma$, the sum contains only finitely many
nonzero terms.
Every term with $m\geq1$ vanishes as $\lambda\to0^+$,
leaving the $m=0$ term $\hbar\Theta(s-\sigma)$.
Consequently,
\begin{align}
 \lim_{\lambda\to0^+}B_{\sigma,\lambda}(s)
 &=\hbar\Theta(s-\sigma),
 \label{eq:coupling-limit}\\
 \lim_{\lambda\to0^+}\ii\hbar P_{\sigma,\lambda}(s)
 &=\hbar\Theta(s-\sigma),
 \qquad s>0,\quad s\neq\sigma\,.
 \label{eq:coupling-limit-poisson}
\end{align}
The limiting value agrees with the free theory for $s>\sigma$,
but on the first interval it gives
\begin{equation}
 \lim_{\lambda\to0^+}B_{\sigma,\lambda}(s)=0
 \quad\neq\quad
 B_{\mathrm{free}}(s)=\hbar,
 \qquad 0<s<\sigma\,.
 \label{eq:free-limit-commutator-mismatch}
\end{equation}
Unlike the local theory in Eq.~\eqref{eq:local-reference-commutator},
the non-local theory therefore retains a discrepancy of $\hbar$ on a
fixed interval of time separations, however small the coupling becomes.
On this interval, the full commutator cannot be written as
$B_{\mathrm{free}}+O(\lambda)$.
Figure~\ref{fig:zero-limit} illustrates this discontinuity.

The origin of this discrepancy can be seen in the characteristic modes. 
For every $\lambda,\sigma>0$, the characteristic equation~\eqref{eq:lambert} has infinitely many roots,
each of which contributes an independent mode pair to the symplectic
two-form~\eqref{eq:omega-modes}. At $\lambda=0$, only the single pair in
Eq.~\eqref{eq:strict-free} remains. 
To identify the contribution responsible for this difference,
we decompose $P_{\sigma,\lambda}$ into the contribution of
the principal root $k_0$ and the sum over all other roots:
\begin{equation}
 P_{\sigma,\lambda}^{(0)}(s)
 :=\frac{\ee^{\ii k_0(s-\sigma)}}{\ii(1+\ii k_0\sigma)},
 \qquad
 R_{\sigma,\lambda}(s)
 :=P_{\sigma,\lambda}(s)-P_{\sigma,\lambda}^{(0)}(s)\,.
 \label{eq:principal-and-remainder}
\end{equation}
Since the bracket $P_{\sigma,\lambda}$ vanishes for $0<s<\sigma$, we have
$R=-P^{(0)}$ there. Consequently,
\begin{equation}
 \lim_{\lambda\to0^+}\ii\hbar
 P_{\sigma,\lambda}^{(0)}(s)=\hbar,
 \qquad
 \lim_{\lambda\to0^+}\ii\hbar
 R_{\sigma,\lambda}(s)=-\hbar,
 \qquad 0<s<\sigma\,.
 \label{eq:coupling-mode-cancellation}
\end{equation}
The principal mode approaches the free commutator, but the remaining modes retain a finite total contribution that cancels it. 
These limits are obtained after evaluating the prescribed
mode sum; they do not assume an interchange of the sum with the limit of each mode. 
Setting $\lambda=0$ at the outset removes these additional directions before the symplectic form is inverted and therefore omits their contribution. 
This explains why an arbitrarily small coupling does
not justify replacing the full canonical structure by the free one.

\subsection{Zero-delay limit at fixed separation}
We next take the zero-delay limit, keeping $\lambda>0$ and the physical time separation $s>0$ fixed. 
The terms in Eq.~\eqref{eq:B-closed-form} are bounded in absolute value by $\hbar(\lambda s)^m/m!$,
independently of $\sigma$. This bound justifies taking the limit term by term in that expansion and gives
\begin{equation}
 \lim_{\sigma\to0^+}B_{\sigma,\lambda}(s)
 =\hbar\sum_{m=0}^{\infty}\frac{(\ii\lambda s)^m}{m!}
 =\hbar\ee^{\ii\lambda s}
 =B_{\mathrm{loc},\lambda}(s),
 \qquad s>0\,.
 \label{eq:delay-limit}
\end{equation}
Thus, at any fixed positive separation, taking the delay to zero agrees with constructing the commutator after setting $\sigma=0$ in the action.

\paragraph{Nonuniform convergence near equal time.} 
Agreement at fixed $s>0$ does not imply uniform convergence near equal time. 
For every $\sigma>0$, the full commutator vanishes on
$0<s<\sigma$, whereas $|B_{\mathrm{loc},\lambda}(s)|=\hbar$ there.
The discrepancy thus has magnitude $\hbar$ on an interval whose width shrinks to zero. 
This is reflected in the two orders of taking the zero-delay and equal-time limits. 
At fixed $\lambda>0$,
\begin{equation}
  \lim_{s\to0^+}\lim_{\sigma\to0^+}
 B_{\sigma,\lambda}(s)=\hbar\,,
 \qquad
 \lim_{\sigma\to0^+}\lim_{s\to0^+}
 B_{\sigma,\lambda}(s)=0\,.
 \label{eq:noncommuting-local-limits}
\end{equation}
Taking $\sigma\to0^+$ first at fixed $s>0$ gives the local commutator;
taking $s\to0^+$ first always probes the interval where $B=0$.
At fixed $s>0$, by contrast, the two orders of taking $\lambda\to0^+$ and
$\sigma\to0^+$ both give $\hbar$.

The two results therefore have distinct meanings. In the zero-coupling limit at fixed $\sigma>0$, the discrepancy from the strictly free theory is confined to $0<s<\sigma$, an interval of fixed width. 
In the zero-delay limit at fixed $\lambda>0$, the difference from the local commutator vanishes at every fixed $s>0$, but a discrepancy of magnitude $\hbar$ remains on the shrinking interval $0<s<\sigma$, where $B=0$ for every $\lambda,\sigma>0$. 
This is the nonuniform convergence near equal time.

\subsection{Remarks}
These results do not undermine the canonical structure familiar from
local theories, but they require care in how the local theory is
approached.
\begin{itemize}
\item
\emph{The jump occurs at shifted equal time.}
For every $\sigma,\lambda>0$, the commutator jumps by $\hbar$. In the
original time labels the jump occurs at $s=\sigma$, which is equal time
in the variables~\eqref{eq:local-kinetic-variables} with local kinetic
term. 
\item
\emph{The local theory is recovered by removing the range of the
non-locality.}
The shift $\sigma$ sets the time range of the non-local term and the
time scales of the additional modes. Sending $\sigma\to0^+$ at fixed
$\lambda$ pushes these modes to arbitrarily short time scales, and the
local commutator $\hbar\ee^{\ii\lambda s}$ is recovered at every fixed
$s>0$, Eq.~\eqref{eq:delay-limit}.
\item
\emph{The order of limits matters.}
The usual equal-time CCR is obtained by taking $\sigma\to0^+$ before
$s\to0^+$. In the opposite order, $s\to0^+$ at finite $\sigma$ probes
the interval before the jump and gives $0$,
Eq.~\eqref{eq:noncommuting-local-limits}.
\item
\emph{Weak coupling is not the same as short range.}
At fixed $\sigma>0$, taking $\lambda\to0^+$ does not commute with
constructing the canonical structure, and on $0<s<\sigma$ the
commutator remains $0$ instead of the free value $\hbar$. A small
coupling alone therefore does not justify using the free CCR: for the commutator family considered here, the expansion in $\lambda$
has zeroth-order term $B_0(s)=\hbar\Theta(s-\sigma)$.
This agrees with the failure of canonical perturbation theory based on free operators noted in Ref.~\cite{Towards}.
\end{itemize}

\section{Summary and discussion}
\label{sec:discussion}
In theories with non-locality in time, the symplectic form
can involve field variations over a time interval, and its
relation to quantum commutators requires careful examination.
Motivated by this issue in SFT, we have studied the finite-shift
model of Ref.~\cite{Towards} to compare its classical and
quantum canonical structures and examine their zero-coupling
and zero-delay limits.
For $b=1$ and $\lambda>0$, this model has infinitely many
characteristic modes while allowing the classical Poisson
bracket and the quantum commutator expectation value to be
calculated independently.
With the prescribed classical mode sum, these calculations give
$B_{\sigma,\lambda}(s)=\ii\hbar P_{\sigma,\lambda}(s)$ for $s>0$,
under the evolution and continuity conditions stated in
Section~\ref{subsec:SD-kernel}.
This agreement provides an explicit consistency check between
the CPS construction and the quantum SD relation for an action
non-local in time.

For every $\lambda>0$, the commutator $B$ vanishes on $0<s<\sigma$, even
though the strictly free theory has $B_{\mathrm{free}}=\hbar$.
The commutator $B$ jumps by $\hbar$ at $s=\sigma$, which is
equal time for the shifted variables with a local kinetic term.
Decomposing the evaluated mode sum shows that, on $0<s<\sigma$, the
additional modes cancel the principal mode even as $\lambda\to0^+$.
Their finite total contribution accounts for the singular zero-coupling limit.
Sending the time shift to zero instead recovers the local commutator at
fixed coupling and fixed positive separation, with nonuniform convergence
near equal time.

The classical bracket is obtained by inverting the symplectic form on
finite-mode subspaces and taking the symmetric mode sum specified in
Appendix~\ref{app:spectral}.
On the quantum side, the BV identity~\eqref{eq:Q-exact-SD}
shows that an EOM insertion and its quantum contact term form
a single $Q_\hbar$-exact combination~\cite{FredenhagenRejzner,MasudaMatsunaga,MatsunagaBVCCR}.
In an operator realization, the EOM and commutator contributions
have to therefore satisfy the SD relation together.
In a local theory, the commutator contributes only at equal time and is
thus separated from the EOM insertion, Eq.~\eqref{eq:local-EOM-CCR-balance}.
In the non-local model, the derivative and delayed terms involve
different operator orderings on $0<s<\sigma$, so the commutator
contributes on this whole interval.
There the SD relation fixes only the sum of the two contributions,
and the joint role of the EOM and the commutator becomes manifest.

The positive-time commutator~\eqref{eq:B-closed-form} agrees with the
constant-background specialization of the construction in
Ref.~\cite{Towards}, up to the chosen value at the jump.
As explained in Section~\ref{subsec:SD-kernel}, its real-$\sigma$ vacuum
prescription realizes our assumptions at the level of formal perturbative
correlators. 
Our analysis complements this quantum construction with an
independent derivation of the classical bracket from the action
and a quantitative account of the singular zero-coupling limit.

The commutator admits an expansion in $\lambda$ with zeroth-order term
$B_0(s)=\hbar\Theta(s-\sigma)$, as in Section~\ref{subsec:SD-kernel}.
This coefficient is the $\lambda\to0^+$ limit of the full commutator
and differs from the strictly free commutator $\hbar$.
Other approaches reduce higher- or infinite-order equations to second-order ones under suitable conditions~\cite{JaenLlosaMolina} or select solutions through perturbative constraints~\cite{Simon}. 
Here all characteristic modes are retained in the symmetric sum. 
Keeping only the principal mode would give
$P_{\sigma,\lambda}^{(0)}$ in Eq.~\eqref{eq:principal-and-remainder} and would omit precisely the contribution responsible for the cancellation.

The correspondence established here is restricted to $s>0$.
Appendix~\ref{app:distributional-extension} constructs a distributional extension to nonpositive separation by additionally requiring the homogeneous delay equation to hold on the whole real line.
Its correspondence with the classical characteristic-mode sum
has not been established here.

Within these limitations, this model makes it possible to follow explicitly both the
collective mode contributions to the classical bracket and the
EOM and commutator contributions to the quantum SD relation.
We conclude with several further observations and open questions:
\vspace{-2mm}
\begin{itemize}
\item
An extension of the classical--quantum correspondence to fluctuations
around different classical solutions when nonlinear interactions are
included. For fixed $b(t)$, the symplectic form of the present quadratic
model is independent of the chosen classical solution.
\vspace{-3mm}
\item
An extension of the analysis to the non-local CPS constructions used in
SFT~\cite{BernardesErlerFirat,BEFPartI,BEFPartII} and to
string-field-theory operators such as $\ee^{-c\partial_t^2}$, $c>0$,
with a suitable domain and time-ordering prescription.
This would test whether the same classical--quantum correspondence holds
there and under what conditions on the correlators.
The backward heat operator $\ee^{-c\partial_t^2}$ is the temporal part of
$\ee^{-c k^2}$ with the mostly-plus Lorentzian metric; it is not a
real-time Gaussian convolution.
\vspace{-3mm}
\item
A Hilbert-space realization with a specified adjoint operation and a positive inner product. Independently of positivity, the jump of the correlator $\avg{\hat a(s)\hat a^\dagger(0)}$ at $s=\sigma$ excludes weak time continuity of $\hat a(s)$ on a common domain containing the states defining this matrix element.
\end{itemize}

\bigskip

\noindent
{\normalfont \bfseries \large Acknowledgments}

Y.~G. would like to thank all members of the particle theory group at
Kyoto University for helpful discussions. 
The work of H. M. is supported by JSPS KAKENHI grant number JP22K14038 and the Asahipen Hikari Foundation.
The work of J.~T. was supported by JSPS KAKENHI Grant Number JP22H05115.
The authors used GPT-6 Astra to assist in developing the contour-integral argument of Appendix~\ref{app:spectral}. All
derivations were checked independently by the authors, who take full
responsibility for the content of this paper.

\medskip

\appendix
\section{The positive-time mode sum}
\label{app:spectral}

In this appendix, we define the field bracket used in
Section~\ref{sec:poisson} through finite-mode restrictions of the
symplectic form and evaluate the resulting symmetric sum.
Throughout this appendix, $\lambda,\sigma>0$ are fixed.

\paragraph{Finite-mode truncation and summation.}
We retain only the mode pairs with $-N\leq n\leq N$ in
Eq.~\eqref{eq:modes}.
Substituting these finite sums into Eq.~\eqref{eq:omega} gives
\begin{equation}
 \Om_N=
 \sum_{n=-N}^{N}C_n\delta\beta_n\wedge\delta\alpha_n,
 \qquad
 C_n=\ii\ee^{\ii k_n\sigma}(1+\ii k_n\sigma)\,.
 \label{eq:omega-finite}
\end{equation}
The condition $C_n=0$ would require $k_n=\ii/\sigma$ and,
by the characteristic equation,
$\lambda=\ii/(\ee\sigma)$.
Thus every $C_n$ is nonzero for real positive $\lambda$ and $\sigma$.
Consequently, $\Om_N$ is nondegenerate, and its inverse gives
Eq.~\eqref{eq:mode-bracket} for $|n|,|m|\leq N$.

Writing $a_N$ and $\ad_N$ for the corresponding truncated fields,
we obtain
\begin{equation}
 P_N(s):=\PB{a_N(t)}{\ad_N(t')}_N
 =\sum_{n=-N}^{N}
 \frac{\ee^{\ii k_n(s-\sigma)}}{\ii(1+\ii k_n\sigma)},
 \qquad s=t-t'\,.
 \label{eq:root-sum-cutoff}
\end{equation}
We define the positive-time bracket kernel by
\begin{equation}
 P_{\sigma,\lambda}(s):=\lim_{N\to\infty}P_N(s),
 \qquad s>0\,.
 \label{eq:root-sum-limit}
\end{equation}
The notation $\sum^{\mathrm{sym}}$ in Eq.~\eqref{eq:root-sum}
denotes this prescription.
For fixed $s>0$, the contour calculation below establishes the
existence of the limit and evaluates it explicitly.

\paragraph{Asymptotic behavior of the roots.}
Taking the modulus and phase of the characteristic equation
$k_n=\lambda\ee^{-\ii\sigma k_n}$ gives
\begin{equation}
 \sigma\operatorname{Re}k_n+\arg k_n\in2\pi\mathbb Z,
 \qquad
 \operatorname{Im}k_n=\frac1\sigma\log\frac{|k_n|}{\lambda}\,.
 \label{eq:k-modulus-phase}
\end{equation}
To fix the branch labels and control the error, we use
Eq.~\eqref{eq:lambert} and the fixed-$z$ expansion
$W_n(z)=2\pi\ii n-\log(2\pi\ii n)+\log z
+O(\log|n|/|n|)$ as $|n|\to\infty$, with principal
logarithms~\cite{Cohen}.
At $z=\ii\lambda\sigma$, this gives, as $n\to+\infty$,
\begin{align}
 k_n&=\frac{2\pi n}{\sigma}
 +\frac{\ii}{\sigma}\log\frac{2\pi n}{\lambda\sigma}+o(1),
 \label{eq:k-large-positive}\\
 k_{-n}&=\frac{-2\pi n+\pi}{\sigma}
 +\frac{\ii}{\sigma}\log\frac{2\pi n}{\lambda\sigma}+o(1)\,.
 \label{eq:k-large-negative}
\end{align}
For fixed $\lambda,\sigma>0$, both remainders are in fact
$O(\log n/n)$. In particular, the $+\pi$ in the real part of
$k_{-n}$ follows from $\arg(-2\pi\ii n)=-\pi/2$.

\paragraph{A contour enclosing the mode sum.}
Define
\begin{equation}
 D(k)=k-\lambda\ee^{-\ii\sigma k},
 \qquad
 F_s(k)=\frac{\ee^{\ii k(s-\sigma)}}{D(k)}\,.
 \label{eq:k-contour-integrand}
\end{equation}
Since $D'(k_n)=1+\ii\sigma k_n\neq0$, the poles of $F_s$ are simple.
Their residues are $\ii$ times the corresponding terms in
Eq.~\eqref{eq:root-sum-cutoff}.

Choose $\gamma>0$ such that $\lambda\ee^{-\sigma\gamma}<\gamma$.
For $\operatorname{Im}k\leq-\gamma$,
$|k|\geq\gamma>\lambda\ee^{-\sigma\gamma}\geq
\lambda|\ee^{-\ii\sigma k}|$, so there are no roots in this region.
Let
\begin{equation}
 L_N=\frac{2\pi N+\pi/2}{\sigma},
 \qquad
 H_N=\frac{M}{\sigma}\log(2\pi N+\pi/2),
 \qquad M>\max\!\left(1,\frac{\sigma}{s}\right)\,,
 \label{eq:k-rectangle-size}
\end{equation}
and let $\Gamma_N$ be the counterclockwise rectangular contour with
vertices $\pm L_N-\ii\gamma$ and $\pm L_N+\ii H_N$.

\begin{figure}[!t]
 \centering
 \includegraphics[width=0.80\linewidth]{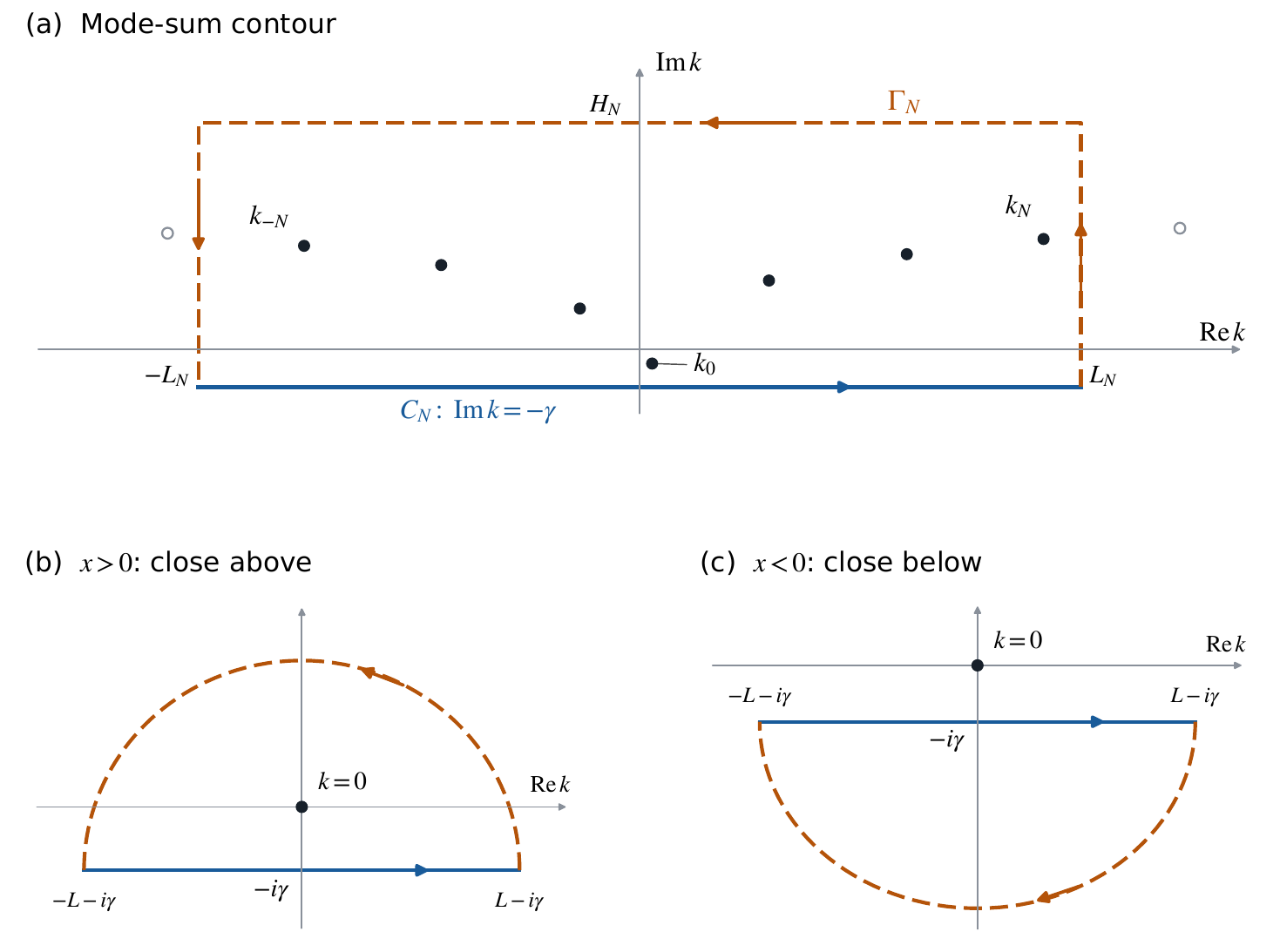}
 \caption{Contours in the complex $k$ plane.
 (a) The counterclockwise rectangle $\Gamma_N$ used to evaluate the mode
 sum. Its blue lower edge runs from left to right; the contributions
 from the three orange edges vanish as $N\to\infty$. Filled dots denote
 enclosed roots, and open dots denote roots outside the contour. The
 root configuration shown is for $\lambda\sigma=1$, $N=3$,
 $\gamma\sigma=1$, and $M=2$.
 (b), (c) Contours for the auxiliary integral $I_m(x)$. For $x>0$, the
 contour closes above and encloses the pole at the origin. For $x<0$,
 it closes below and encloses no poles.}
 \label{fig:appendix-contours}
\end{figure}

For sufficiently large $N$, the vanishing real-part errors in
Eqs.~\eqref{eq:k-large-positive} and \eqref{eq:k-large-negative} place
$-L_N$ between $\operatorname{Re}k_{-(N+1)}$ and
$\operatorname{Re}k_{-N}$, and $L_N$ between
$\operatorname{Re}k_N$ and $\operatorname{Re}k_{N+1}$.
The logarithmic imaginary-part growth and $M>1$ place all roots with
$|n|\leq N$ below $H_N$, while the lower edge lies below every root.
Thus the contour encloses precisely $k_{-N},\ldots,k_N$;
see Fig.~\ref{fig:appendix-contours}(a). Therefore,
\begin{equation}
 \frac1{2\pi\ii}\oint_{\Gamma_N}F_s(k)\dd k
 =\sum_{n=-N}^{N}\frac{\ee^{\ii k_n(s-\sigma)}}{1+\ii\sigma k_n}
 =\ii P_N(s)\,.
 \label{eq:k-residue-sum}
\end{equation}

On the upper edge, $\lambda\ee^{\sigma H_N}$ dominates $|k|=O(N)$,
so $|F_s(k)|=O(\ee^{-sH_N})$. The upper-edge integral is consequently
\begin{equation}
 O(L_N\ee^{-sH_N})=O(N^{1-Ms/\sigma})\longrightarrow0\,.
 \label{eq:k-top-bound}
\end{equation}
On the vertical edges, $k=\pm L_N+\ii y$, we have
$\ee^{\mp\ii\sigma L_N}=\mp\ii$. For sufficiently large $N$, the
modulus of the denominator is bounded below by a positive constant
times $L_N+\lambda\ee^{\sigma y}$. Splitting the integral at
$y=\sigma^{-1}\log(L_N/\lambda)$, each vertical-edge contribution is
bounded by
\begin{equation}
 O\!\left(
 \int_{-\gamma}^{H_N}
 \frac{\ee^{-(s-\sigma)y}}{L_N+\lambda\ee^{\sigma y}}\dd y
 \right)
 =\begin{cases}
 O(N^{-s/\sigma}),&0<s<\sigma,\\
 O(N^{-1}\log N),&s=\sigma,\\
 O(N^{-1}),&s>\sigma
 \end{cases}\,,
 \label{eq:k-side-bound}
\end{equation}
and hence vanishes. Provided the symmetric limit of the lower-edge
integral exists, Eq.~\eqref{eq:k-residue-sum} yields
\begin{equation}
 P_{\sigma,\lambda}(s)=-\frac1{2\pi}
 \lim_{L\to\infty}\int_{-L-\ii\gamma}^{L-\ii\gamma}
 \frac{\ee^{\ii k(s-\sigma)}}{k-\lambda\ee^{-\ii\sigma k}}\dd k,
 \qquad s>0\,.
 \label{eq:k-line-integral}
\end{equation}
We next establish the existence of this limit and evaluate it.

\paragraph{Termwise integration.}
On the line $\operatorname{Im}k=-\gamma$,
$|\lambda\ee^{-\ii\sigma k}/k|\leq
\rho:=\lambda\ee^{-\sigma\gamma}/\gamma<1$.
We may therefore expand
\begin{equation}
 \frac1{k-\lambda\ee^{-\ii\sigma k}}
 =\frac1k+\sum_{m=1}^{\infty}
 \frac{\lambda^m\ee^{-\ii m\sigma k}}{k^{m+1}}\,.
 \label{eq:k-geometric}
\end{equation}
For the terms with $m\geq1$,
\begin{equation}
 \sum_{m=1}^{\infty}
 \left|\frac{\lambda^m\ee^{-\ii m\sigma k}}{k^{m+1}}\right|
 \leq\frac{\lambda\ee^{-\sigma\gamma}}{1-\rho}\frac1{|k|^2}\,.
 \label{eq:k-integrable-bound}
\end{equation}
The right-hand side is integrable along the line. Moreover,
$|\ee^{\ii k(s-\sigma)}|=\ee^{\gamma(s-\sigma)}$ is constant there,
so the sum and integral may be interchanged for these terms. The
$m=0$ term is treated separately by taking the symmetric limit.

The required auxiliary integral is
\begin{equation}
 I_m(x)=\frac1{2\pi\ii}\lim_{L\to\infty}
 \int_{-L-\ii\gamma}^{L-\ii\gamma}
 \frac{\ee^{\ii kx}}{k^{m+1}}\dd k,
 \qquad m=0,1,\ldots\,.
 \label{eq:k-basic-integral}
\end{equation}
For $x>0$, the exponential decays in the upper half-plane, and we close
the contour with an upper semicircle. The semicircle contribution
vanishes, while the residue at the origin is $(\ii x)^m/m!$.
For $x<0$, we close below; there are no poles in that region, so the
integral vanishes. The contours are shown in
Fig.~\ref{fig:appendix-contours}(b), (c). Thus,
\begin{equation}
 I_m(x)=\frac{(\ii x)^m}{m!}\Theta(x)\,.
 \label{eq:k-basic-value}
\end{equation}
At $x=0$, direct integration gives
$I_0(0)=\lim_{L\to\infty}\pi^{-1}\arctan(L/\gamma)=1/2$,
whereas $I_m(0)=0$ for $m\geq1$. Accordingly, we use $\Theta(0)=1/2$
in Eq.~\eqref{eq:k-basic-value} and understand its $m=0$ case as
$I_0(x)=\Theta(x)$.

Substituting Eq.~\eqref{eq:k-geometric} into
Eq.~\eqref{eq:k-line-integral} and using
Eq.~\eqref{eq:k-basic-value} with $x=s-(m+1)\sigma$ gives
\begin{equation}
 P_{\sigma,\lambda}(s)=-\ii\sum_{m=0}^{\infty}
 \frac{(\ii\lambda)^m}{m!}
 [s-(m+1)\sigma]^m\Theta(s-(m+1)\sigma),
 \qquad s>0\,.
 \label{eq:k-sum-result}
\end{equation}
This reproduces Eq.~\eqref{eq:P-positive}. At fixed $s>0$, only finitely
many terms contribute. In particular, $P_{\sigma,\lambda}(\sigma)=-\ii/2$,
which differs from the right-hand limit
$P_{\sigma,\lambda}(\sigma^+)=-\ii$.

\section{A distributional extension to nonpositive separation}
\label{app:distributional-extension}

For fixed $\lambda,\sigma>0$, we extend the positive-time
kernel~\eqref{eq:B-closed-form} by imposing the homogeneous delay equation
\begin{equation}
 \ii\partial_s B_{\mathrm{ext}}(s)
 +\lambda B_{\mathrm{ext}}(s-\sigma)=0
 \label{eq:B-global-EOM}
\end{equation}
distributionally on the whole real line.
This is an additional condition beyond the SD relation.
Rearranging this equation and shifting the argument gives
\begin{equation}
 B_{\mathrm{ext}}(s)
 =-\frac{\ii}{\lambda}\partial_s B_{\mathrm{ext}}(s+\sigma).
 \label{eq:B-backward-recursion}
\end{equation}
Thus, knowing the kernel at later separation allows us to extend it
backwards in steps of $\sigma$.

The positive-time kernel satisfies
$B_{\sigma,\lambda}(s)=0$ for $0<s<\sigma$ and $B_{\sigma,\lambda}(s)=\hbar$ for
$\sigma<s<2\sigma$.
Its distributional derivative on $0<s<2\sigma$ is therefore
$\hbar\delta(s-\sigma)$.
Equation~\eqref{eq:B-backward-recursion} then gives
\begin{equation}
 B_{\mathrm{ext}}(s)
 =-\frac{\ii\hbar}{\lambda}\delta(s),
 \qquad -\sigma<s<\sigma.
 \label{eq:B-extension-first-step}
\end{equation}
The jump at $s=\sigma$ thus requires a delta function at $s=0$.
Applying the same recursion twice more gives
\begin{align*}
 B_{\mathrm{ext}}(s)
 &=-\frac{\hbar}{\lambda^2}\delta'(s+\sigma),
 &&-2\sigma<s<0,\\
 B_{\mathrm{ext}}(s)
 &=\frac{\ii\hbar}{\lambda^3}\delta''(s+2\sigma),
 &&-3\sigma<s<-\sigma.
\end{align*}
Each step shifts the support to the left by $\sigma$, adds one
derivative, and multiplies the coefficient by $-\ii/\lambda$.
The backward construction therefore produces successive inverse
powers of $\lambda$.

Let $B_{\mathrm R}$ denote Eq.~\eqref{eq:B-closed-form} extended
by zero to $s\leq0$.
Continuing the recursion yields
\begin{equation}
 B_{\mathrm{ext}}(s)
 =B_{\mathrm R}(s)
 +\hbar\sum_{n=0}^{\infty}
 \left(-\frac{\ii}{\lambda}\right)^{n+1}
 \delta^{(n)}(s+n\sigma).
 \label{eq:B-distributional-extension}
\end{equation}
Here $\delta^{(n)}$ denotes the $n$th distributional derivative
of the Dirac delta.
The added terms are supported at $0,-\sigma,-2\sigma,\ldots$,
so they leave the positive-time kernel unchanged.
Every compact interval meets only finitely many of these support
points. The sum is therefore locally finite and defines a distribution
for each fixed $\lambda,\sigma>0$.

This extension contains delta-function terms and lies outside the
bounded-correlator assumptions of Section~\ref{subsec:SD-kernel}.
We have not established whether it agrees with a distributional
limit of the classical characteristic-mode sum.

\end{document}